\documentclass[twocolumn,aps,pra,nofootinbib,superscriptaddress,amsmath,amssymb,longbibliography]{revtex4-1}
\usepackage{graphicx}
\pdfoutput=1
\usepackage{bm}
\usepackage{soul}
\usepackage[english]{babel}
\usepackage[colorlinks,citecolor=blue,urlcolor=blue,linkcolor=blue]{hyperref}
\usepackage{float}

\usepackage{epsfig}
\usepackage{amsmath}
\usepackage{amssymb}
\usepackage{graphicx}

\begin{document}
\title{Rydberg electromagnetically induced transparency and absorption of strontium triplet states in a weak microwave field}
\author{Yan-Li Zhou}\email{ylzhou@nudt.edu.cn}
\affiliation{Department of Physics, College of Liberal Arts and Sciences, and Interdisciplinary Center for Quantum Information, National University of Defense Technology, Changsha 410073, China}
\author{Dong Yan}
\affiliation{School of Science and Key Laboratory of Materials Design and Quantum Simulation, Changchun University, Changchun 130022, China}
\author{Weibin Li}\email{weibin.li@nottingham.ac.uk}
\affiliation{School of Physics and Astronomy, and Centre for the Mathematics and Theoretical Physics of Quantum Non-equilibrium Systems, University of Nottingham, Nottingham NG7 2RD, United Kingdom}

\begin{abstract}
We study theoretically laser excitation of Rydberg triplet states of strontium atoms in the presence of weak microwave (MW) fields. Starting from the ground state $5s^2\,^1S_0$, the Rydberg excitation is realized through the metastable, triplet $5s5p\,^3P_1$ state, whose decay rate $\gamma_2$ is $2\pi\times 7.5$ kHz, much smaller than the one in the singlet state or alkali atoms. The influences of $\gamma_2$ on the transparency and absorption spectrum in the electromagnetically induced transparency (EIT), and electromagnetically induced absorption (EIA) regime are examined. Narrow transparent windows (EIT) or absorption peaks (EIA) are found, whose distance in the spectrum depends on the Rabi frequency of the weak MW field. It is found that the spectrum exhibits higher contrast than using the singlet state or alkali atoms in typical situations. Using the metastable intermediate state, we find that resonance fluorescence of Sr gases exhibits very narrow peaks, which are modulated by the MW field. When the MW field is weaker than the probe and control light, the spectrum distance of these peaks are linearly proportional to $\Omega_m$. This allows us to propose a new way to sense very weak MW fields through resonance fluorescence. Our study shows that the Sr triplet state could be used to develop the Rydberg MW electrometry that gains unique advantages.
\end{abstract}

\maketitle

\section{Introduction}

Using their large electric-dipole moments and many dipole allowed transitions, Rydberg atoms are utilized to sense weak electromagnetic fields from the microwave (MW) to terahertz frequency range~\cite{Degen2017,Wade2017,Adams2020,Sedlacek2012}. These sensing applications are largely based on alkali atoms, such as Cs \cite{Kumar2017, Kumar2017ScienceReport,Jing2020} and Rb ~\cite{Sedlacek2012, Gordon2014}. Commonly, atoms are laser excited to a low-lying, intermediate state, and then to a Rydberg state by a control light. A MW field couples the Rydberg state near resonantly to a different Rydberg state. Both Rydberg states have long lifetime ($10-100\, \mu$s typically), while the linewidth of the intermediate state is large~\cite{millen_two-electron_2010,Ding2018, Qiao2019} (e.g. about $2\pi \times 6.07$ MHz in the $5p$ state of Rb and $2\pi\times 5.22$ MHz in the $6p$ state of Cs). To sensitively probe the electric component of  MW fields, one typically relies on two quantum interference phenomena, i.e., electromagnetically induced transparency (EIT) and the Autler-Townes (AT) effects \cite{Sedlacek2012, Holloway2014APL,Fan2015,Jing2020}. Here the Rydberg-MW field coupling gives rise to two Rydberg dressed states, which generate two transparent window (the AT splitting). The distance of the AT splitting is determined by the Rabi frequency of the MW electric field, which serves the basis in the Rydberg atom based MW electrometry. This method is influenced by various parameters, such as the control laser strength and decay rate in the intermediate state. Especially the large decay rate  affects the lowest MW field that can be sensed~\cite{Liao2020}. A different method, based on the electromagnetically induced absorption (EIA)~\cite{akulshin_electromagnetically_1998}, has been demonstrated experimentally~\cite{Liao2020}. In the EIA method, the intermediate state is adiabatically eliminated through large single photon detuning.  The MW coupled AT states lead to two absorption maxima, whose distance equals to the MW Rabi frequency when the latter is much larger than the AC Stark shift.  This approach largely avoids impacts from the intermediate state, and can sense weak MW fields as low as $100\,\mu$V/cm~\cite{Liao2020}. However the visibility of the absorption spectrum  is suppressed, due to that the overall absorption is much weaker.
	
In recent years, Rydberg states of alkaline-earth atoms have attracted a growing interest~\cite{Dunning2016, Madjarov2020}. Experiments have observed the Rydberg state excitation~\cite{millen_two-electron_2010}, blockade effects~\cite{desalvo_rydberg-blockade_2016,qiao_rydberg_2021}, Rydberg dressing~\cite{gaul_resonant_2016}, and Rydberg polaron~\cite{camargo_creation_2018} in gases of strontium atoms. Compared to alkali atoms, one of the key differences is that Rydberg states can be excited through intermediate, triplet states. For example, the linewidth of triplet $5s5p\,^3P_1$ state of Sr atoms is $\gamma_2=2\pi \times7.5$ kHz. The long coherence time in the triplet state plays important roles in the study of Rydberg physics of alkaline-earth atoms when excited through EIT~\cite{Dunning2016}, such as quantum clock networks \cite{komar2014}, spin squeezing~\cite{Gil2014} via Rydberg dressed interactions~\cite{robicheaux_calculations_2019,tan_dynamics_2021} and Rydberg atom based high-precision quantum metrology~\cite{Gil2014,Kessler2014, Kaubruegger2019}. When developing MW sensing with alkaline-earth atoms, the employment of the singlet or triplet intermediate state will have different optical response~\cite{couturier_measurement_2019,hu_narrow-line_2021}.  Therefore it is interesting to understand roles played by the triplet state in EIT as well as in developing MW electrometry with Sr atoms.

\begin{figure*}
	\includegraphics[scale=0.8]{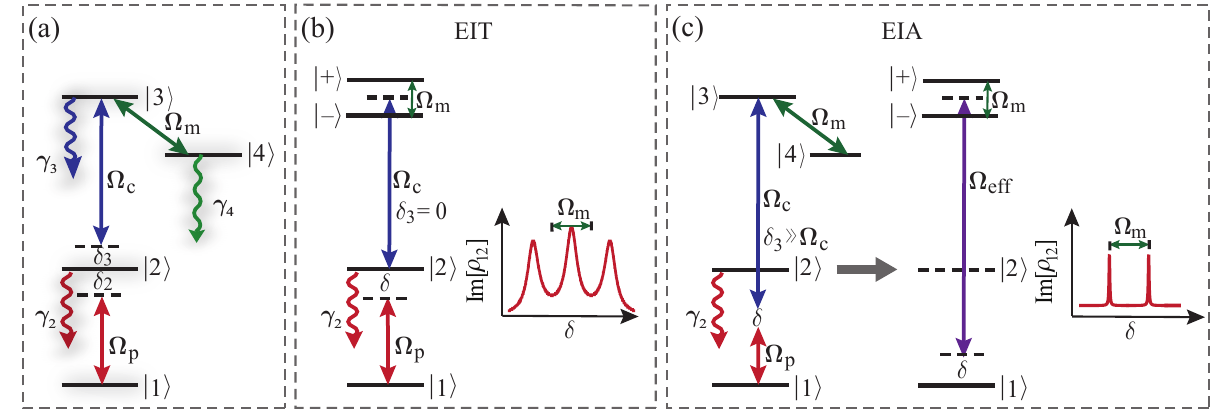}
	\caption{(a) Level scheme. A probe laser  (Rabi frequency $\Omega_p$ and detuning $\delta_2$) and a control laser (Rabi frequency $\Omega_{c}$ and detuning $\delta_3$) couple the transitions between state $|1\rangle\to|2\rangle$, and $|2\rangle\to |3\rangle$, respectively. A near resonant MW field couples Rydberg states $|3\rangle\to |4\rangle$ with Rabi frequency $\Omega_{m}$. (b) EIT and AT effect. When the control laser is resonant $\delta_3 = 0$,  the resonant MW field splits the EIT transmission maximum into two peaks. The distance of the two peaks,  i.e. the AT splitting of the dressed states $|+\rangle$ and $|-\rangle$, equals to $\Omega_m$ approximately. (c) EIA scheme is achieved when both the control and probe light are far detuned, while two-photon resonance is maintained approximately. The MW field induces two absorption peaks with splitting $\Omega_{m}$ near the two-photon resonant transition. }
	\label{fig:level}
\end{figure*}

In this work, we study AT effect of Sr atoms when Rydberg states are coupled by a MW field. The Rydberg state is laser excited via the triplet $5s5p\,^3P_1$ state. We focus on influences of the weak decay rate in the metastable state. In the stationary EIT spectrum of Sr atoms, transparent windows are surrounded by strong absorption peaks, leading to high contrast signals even when both the MW and probe field Rabi frequency are in the same order of magnitude of the triplet decay rate $\gamma_2$. Under the large one-photon detuning condition, the distance between EIA peaks can be used to determine the MW Rabi frequency~\cite{Liao2020}. We show that high contrast can be realized through EIA when both the MW and probe field are close to $\gamma_2$.

Furthermore, we propose a new way to sense MW field based on resonance fluorescence \cite{PhysRev.188.1969}. Resonance fluorescence is widely used in optical and atomic physics~\cite{scully_quantum_1997}. Different from the EIT and EIA method, photons randomly scattered by atoms are measured in resonance fluorescence. We show the narrow linewidth of Sr atoms leads to well-separated fluorescence peaks. We identify the linear regime where the distance of the peaks are proportional to the MW Rabi frequency. Moreover, the time-dependent resonance fluorescence is employed to capture the slow time scale in the optical response, determined by $\gamma_2$. To the best of our knowledge, this is the first time to show that resonance fluorescence can be used to sense MW fields with Rydberg states of Sr atoms.

The remainder of the paper is organized as follows. In Sec. II, we present the system whose dynamics is modeled by a quantum master equation of four-level atoms. In Sec. III, optical responses of the system in the presence of MW fields are discussed. We show that the spectrum of the optical response in both the EIT and EIA regime. A comparison between Rb and Sr atoms shows that high contrast optical responses can be obtained in Sr atoms. In sec. IV, we investigate roles of the intermediate state played in the dynamics and resonance fluorescence. It is found that resonance fluorescence can also be used to quantify the MW field.  The main results and conclusions of this work are summarized in Sec. V.

\section{System and model}
The laser-MW-coupled atom is described by a four-level model, as depicted in Fig.~\ref{fig:level}(a). The probe light of frequency $\omega_p$ drives the transition from the atomic ground state $|1\rangle$ to an intermediate state $|2\rangle$ with Rabi frequency $\Omega_{p}$ and detuning $\delta_2=\omega_{21}-\omega_p$. A strong control light of frequency $\omega_c$ acts on the transition between state $|2\rangle$ and Rydberg state $|3\rangle$ with Rabi frequency $\Omega_{c}$ and detuning $\delta_3=\omega_{32}-\omega_c$. The MW field with frequency $\omega_m$ couples two Rydberg states $|3\rangle$ and $|4\rangle$ with Rabi frequency $\Omega_{m}$ and detuning $\delta_m=\omega_{43}-\omega_m$.  $\gamma_{i}$ denotes the decay rate from state $|i\rangle$ to $|1\rangle$. In the electric dipole and rotating-wave  approximation, Hamiltonian of the system is given by ($\hbar \equiv 1$) \cite{Liao2020}:
\begin{equation}
    H = H_d +\frac{1}{2}\left[\Omega_\mathrm{p} \sigma_{12}+ \Omega_\mathrm{c} \sigma_{23} +  \Omega_{\mathrm{m}} \sigma_{34}+ \mathrm{H. c.}\right],
\end{equation}
where $\sigma_{ab} = |a\rangle\langle b|$  ($a,b=$1, 2, 3 and 4), and $H_d=\delta_2 \sigma_{22}+ \delta \sigma_{33} + \delta_4 \sigma_{44}$ with $\delta = \delta_2 + \delta_3$ and $\delta_4 = \delta_2 +\delta_3+ \delta_m$. The two-photon detuning takes into account of the single-photon probe field detuning $\delta_2$ and control field detuning $\delta_3$. Due to spontaneous decays in the intermediate state and Rydberg states, dynamics of the density matrix $\rho$ is governed by a quantum master equation,
\begin{equation}
	\label{eq:ME}
\frac{d}{dt}\rho = \mathcal L \rho = - i [H, \rho] + \sum_{i=2}^4 \gamma_{i}L(\sigma_{1i})\rho,
\end{equation}
with the Liouville $\mathcal{L}$ and dissipator $L(\cdot)\rho = (\cdot) \rho (\cdot)^{\dagger}- \{(\cdot)^{\dagger}(\cdot), \rho\}/2$, and the jump operator is given by $\sigma_{1j} = |1\rangle\langle i|$. Here we assume that Rydberg states directly decay to the ground state $|1\rangle$, while neglecting cascade decay to multiple low-lying electronic states. Dephasing of various electronic states has been neglected.
\begin{figure*}
	\includegraphics[width=0.98\textwidth]{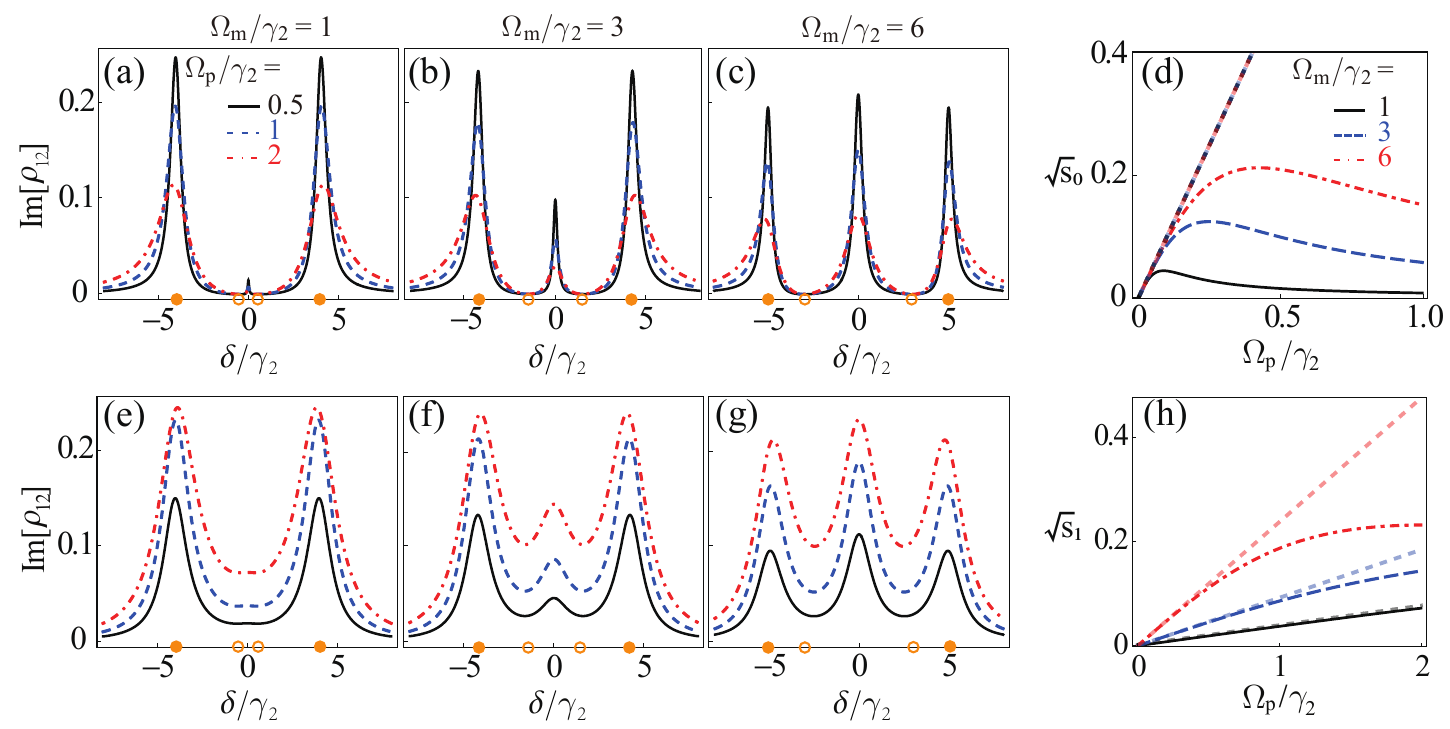}
	\caption{(a-c) Imaginary parts of the coherence Im[$\rho_{12}$] by varying the two-photon detuning for (a) $\Omega_m=\gamma_2$, (b) $\Omega_m= 3\gamma_2$ and (c) $\Omega_m= 6\gamma_2$ when $\gamma_3=\gamma_4=0$. In each panel, the solid, dashed and dot-dashed curves correspond to $\Omega_p= 0.5\gamma_2,\, \gamma_2,\, \text{and}\ 2 \gamma_2$. The open circle and filled circle in the axis give the minimal and maximal locations of Im[$\rho_{12}$] at $\delta=\pm\Omega_m/2$ and $\delta=\pm\sqrt{\Omega_m^2+\Omega_c^2}/2$, respectively. The numerical data agree with the perturbative calculations excellently. (d)  $\sqrt {s_0}$ when $\gamma_3=\gamma_4=\delta_2=0$. The other parameters are $\delta_3=\delta_m=0$, and $\Omega_c=8\gamma_2$. The solid, dashed, and dot-dashed curves correspond to $\Omega_m= \gamma_2, 3\gamma_2, \ \text{and}\  6\gamma_2$. The dotted curves are the analytical results shown in Eq. (9). (e-g) Im$[\rho_{12}]$ for (e) $\Omega_m=\gamma_2$, (f) $\Omega_m= 3\gamma_2$ and (g) $\Omega_m= 6\gamma_2$ when $\gamma_3=\gamma_4=2\gamma_2$. (h) $\sqrt {s_1}$ when  $\gamma_3=\gamma_4=2\gamma_2$. The solid, dashed, and dot-dashed curves correspond to $\Omega_m= \gamma_2, 3\gamma_2, \ \text{and}\  6\gamma_2$, while the dotted curves are the analytical results shown in Eq. (10). }
	\label{fig:absorption}
\end{figure*}

Focusing on Rydberg triplet states of Sr atoms, we consider the ground state $|1\rangle=|5s^2\,^1S_0\rangle$, and intermediate state $|2\rangle=|5s5p\,^3P_1\rangle$. State $|2\rangle$ couples to a Rydberg state $|3\rangle=|5sns\,^3S_1\rangle$~\cite{desalvo_rydberg-blockade_2016,qiao_rydberg_2021}, or $|5snd\,^3D_1\rangle$~\cite{Ding2018}. A MW field couples state $|3\rangle$ to a neighboring Rydberg state $|4\rangle = |5sn'p\,^3P_1\rangle$. Depending on principal quantum numbers, lifetimes in Rydberg states typically range from a few to tens of microseconds~\cite{kunze_lifetime_1993}, leading to decay rates $\gamma_{3,4}\sim 2\pi \times 1\dots 100$ kHz. In the following, we will compare the optical response of Sr triplet states with conventional situations, where the decay rate of the intermediate state is large. The latter can be realized, for example, by using the Sr singlet state or alkali atoms. To be concrete, the $5S_{1/2}(F=3)\to 5P_{3/2}(F=4)\to50D_{5/2}\to51P_{3/2}$ transition of $^{85}\mathrm{Rb}$ atoms will be considered as an example, where $\gamma_2 \approx 2\pi\times 6 $ MHz~\cite{Sadighi-Bonabi2015}. While the Rydberg decay rate is similar in both cases, the order of magnitude difference of the decay rate in the intermediate states leads to very different optical response, as will be illustrated in the following.

\section{MW modulated EIT and EIA}
The transmission of the probe light is $I_f=I_0\exp(-\alpha L)$ where $L$ and $\alpha$ are the length of the medium and absorption coefficient. The absorption coefficient $\alpha\propto \text{Im}(\rho_{12})$~\cite{fleischhauer_electromagnetically_2005}. Hence our analysis will be focusing on $\rho_{12}$. First, the steady state of the master equation can be solved perturbatively~\cite{fleischhauer_electromagnetically_2005}. In the limit of the weak probe light and with the initial state $\rho(0)=|1\rangle\langle 1|$,  the master equation in the linear order of $\Omega_p$ is approximately given by
\begin{eqnarray}\label{subequation}
\dot{\rho_{12}} &&\simeq  (i \delta_2 - \frac{\gamma_2}{2}) \rho_{12} + i\frac{\Omega_c}{2} \rho_{13} + i\frac{\Omega_p}{2},\label{sub1}\\
\dot{\rho_{13}} &&\simeq  (i \delta - \frac{\gamma_3}{2}) \rho_{13} + i\frac{\Omega_c}{2} \rho_{12} + i\frac{\Omega_{m}}{2}\rho_{14},\label{sub2}\\
\dot{\rho_{14}} &&\simeq  (i \delta_4 - \frac{\gamma_4}{2}) \rho_{14} + i\frac{\Omega_{m}}{2}\rho_{13}.
\label{eq:approxME}
\end{eqnarray}
The analytic expression of the coherence between states $|1\rangle$ and $|2\rangle$ in the steady state is~\cite{Gao2019, Sadighi-Bonabi2015},
\begin{equation}
\rho_{12} \approx i\frac{|\Omega_{m}|^2 + \Gamma_{3} \Gamma_{4}}{\Gamma_{2}|\Omega_{m}|^2 + \Gamma_{4}|\Omega_{c}|^2  + \Gamma_{2}\Gamma_{3}\Gamma_{4}}\Omega_p
\end{equation}
where $\Gamma_{2} = \gamma_{2} - 2i\delta_{2}$, $\Gamma_{3} = \gamma_{3} - 2i\delta$, and $\Gamma_{4} = \gamma_{4} - 2i\delta_{4}$. The absorption of the probe light is determined by the imaginary part of $\rho_{12}$, while the real part gives the dispersion. Here a crucial approximation in deriving the steady solution is that $\Omega_p$ is assumed to be small. We note that such perturbative approach is valid in case of alkali atoms (e.g. Rb) where the condition $\gamma_2\gg \Omega_2$ can be met. In case of Sr triplet states, we will examine the validity of the perturbative result by comparing the perturbative and numerical calculations.

In the EIT regime, both the control light and MW field are resonant with respect to the underlying transition, i.e. $\delta_3=\delta_m=0$. The MW field causes the AT splitting, whose distance depends on the MW Rabi frequency. This relation can be used to accurately sense properties of MW fields~\cite{Sedlacek2012}. Recently Liao et al. ~\cite{Liao2020} have experimentally studied EIA of Rb atoms. Their experiment shows that $\Omega_m$ can be determined through measuring the absorption spectrum. EIA of Sr atoms has not been studied thoroughly, which will be examined in the following.

\subsection{EIT regime}
In the case of alkali atoms decay in Rydberg states is typically not important, as typically $\gamma_3\sim \gamma_4\ll \gamma_2$. When the coupling and MW field are resonant with respect to the transitions, i.e. $\delta_3= \delta_m=0$, the coherence is reduced to
\begin{eqnarray}
 \rho_{12}\approx-\frac{4\delta^2-\Omega_m^2}{2\delta(4\delta^2-\Omega_c^2-\Omega_m^2)
 +i\gamma_2(4\delta^2-\Omega_m^2)}\Omega_p.
\end{eqnarray}
Both the imaginary and real parts of the coherence vanish when $\delta=\pm \Omega_m/2$, leading to two transparent windows. When $\delta=0$, or $\delta=\pm \sqrt{\Omega_c^2+\Omega_m^2}/2$, the coherence becomes a pure imaginary number, $\rho_{12}=-i\Omega_p/\gamma_2$. This leads to strong absorption, whose peak increases with smaller $\gamma_2$. Moreover, the coherence is pure imaginary when all the fields are resonant,
\begin{equation}
\rho_{12} = i \frac{\gamma_3\gamma_4 +\Omega_m^2}{\gamma_4\Omega_c^2 +\gamma_2(\gamma_3\gamma_4 +\Omega_m^2)}\Omega_p.
\end{equation}
We define the saturation of the absorption with $s=|\rho_{12}|^2$. With $\gamma_2\gg \gamma_3,\gamma_4$, the saturation is approximately given by
\begin{equation}
	\label{eq:saturation}
s\approx s_0=\frac{\Omega_p^2}{\gamma_2^2},
\end{equation}
which indicates the saturation is small in conventionally EIT when $\gamma_2\gg\Omega_p$.
\begin{figure}[h]
	\includegraphics[width=0.48\textwidth]{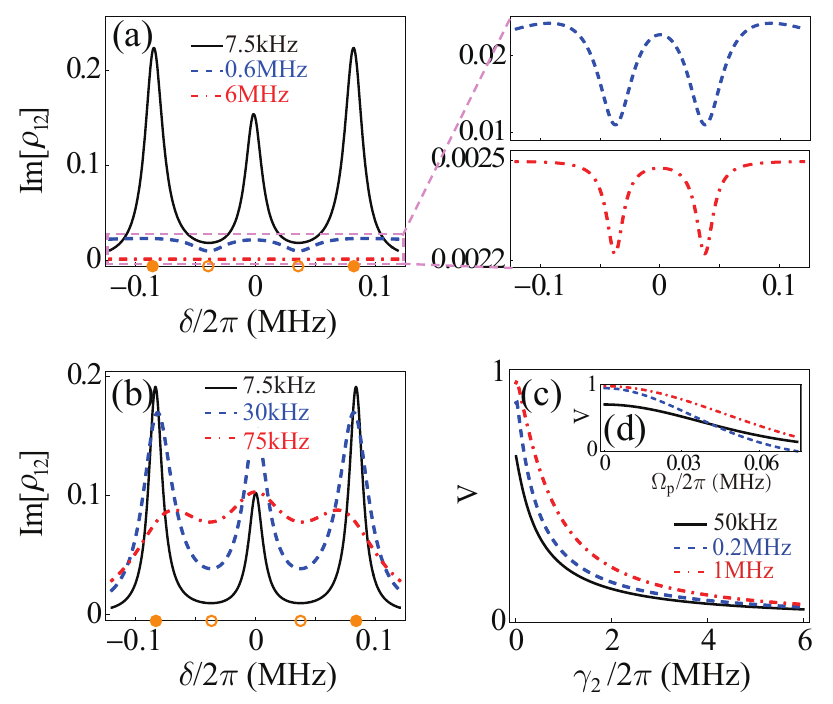}
	\caption{(a) Im[$\rho_{12}$] with $\gamma_2/2\pi = 7.5$ kHz (black solid line), $ 0.6$ MHz (blue dashed line) and $ 6$ MHz (red dot-dashed line). The other parameters are $\delta_3=\delta_m =0$, $\gamma_3/2\pi = \gamma_4/2\pi = 15$ kHz, $\Omega_p = \gamma_3$, $\Omega_c = 10\gamma_3$, $\Omega_m = 5\gamma_3$. Open circle and filled circle in the axis give the minimal and maximal locations of absorption coefficient. The left panel shows a zoom in the absorption. (b) Im[$\rho_{12}$] with $\Omega_p/2\pi = 7.5$ kHz (black solid line), $30$ kHz (blue dashed line) and $ 75$ kHz (red dot-dashed line). Here $\gamma_2/2\pi = 7.5$ kHz and the other parameters are the same with (a). (c) The visibility as function of $\gamma_2$ with $\Omega_m/2\pi = 50$ kHz (black solid line), 0.2 MHz (blue dashed line) and 1 MHz (red dot-dashed line). (d) The visibility as function of $\Omega_p$ with  $\gamma_2/2\pi=7.5$ kHz, $\Omega_m /2\pi = 50$ kHz (black solid line), 0.2 MHz (blue dashed line) and 1 MHz (red dot-dashed line).}
	\label{fig:absorption(exp)}
\end{figure}

To verify the perturbation analysis, we show numerical results  in Figs.~\ref{fig:absorption}(a)-(d) obtained by solving the master equation~(\ref{eq:ME}). Transparent windows are found at $\delta=\pm \Omega_m/2$, denoted by the open circles in Figs.~\ref{fig:absorption}(a)-(c), which agree with the numerical result. Two absorption peaks are visible at $\delta=\pm \sqrt{\Omega_m^2 +\Omega_c^2}/2$, and a weak absorption line appears at $\delta=0$, whose height is determined by $s_0$. This absorption line becomes stronger when increasing $\Omega_m$. In Fig.~\ref{fig:absorption}(d) we plot the resonant absorption at $\delta = 0$. When $\Omega_p$ is much smaller than $\gamma_2$, the saturation $s$ is quadratic with $\Omega_p$, see Eq.~(\ref{eq:saturation}). When $\Omega_p$ is large, the perturbation theory can not predict the absorption accurately.

The above analysis shows cares must be taken when considering Sr atoms, where one typically encounters  $\gamma_{2} < \Omega_p$. Moreover, as Rydberg state decay rates $\gamma_{3}$ and $
\gamma_4$ are comparable with the respective $\gamma_2$, according Eq. (6) one can get $\text{Im}[\rho_{12}] \propto \gamma_3\gamma_4$ at the transparent windows centered at $\delta=\pm\Omega_m/2$. It means that the minima of $\text{Im}[\rho_{12}]$ does not reach zero with non-zero $\gamma_3$ and $\gamma_4$ (see Fig.~\ref{fig:absorption}(e)-(g)). Therefore the probe light will experience losses, regardless of the detuning, and this could reduce the visibility of the transmission spectrum.

With non-zero  $\gamma_3$ and $\gamma_4$, the saturation at resonance becomes
\begin{equation}
\label{eq:saturationRydberg}
s_{1} \approx \left[\frac{(\gamma_{3}^2 + \Omega_m ^2)\Omega_p}{\gamma_2(\Omega_m^2+\gamma_3^2)+\gamma_3\Omega_c^2}\right]^2,
\end{equation}
where we have assumed $\gamma_3 = \gamma_4$. Eq.~(\ref{eq:saturationRydberg}) indicates that the saturation will be affected by the finite Rydberg state decay rates $\gamma_{3(4)}$,  $\Omega_p$ and $\Omega_m$, even when $\Omega_p \ll \gamma_2$ (see Fig.~\ref{fig:absorption}(h)). In addition to the absorption peak at resonance, two more absorption peaks are found at $\delta\approx \pm \sqrt{\Omega_c^2+\Omega_m^2}/2$, which can be seen in Figs.~\ref{fig:absorption}(e)-(g).

The distance between the AT splitting is independent of $\gamma_2$ and $\Omega_p$, as shown in Fig.\ref{fig:absorption(exp)}(a-b). A notable change is that the absorption becomes stronger and its linewidth becomes smaller when decreasing $\gamma_2$.  When fixing $\gamma_2$, larger $\Omega_p$ will decrease the absorption and increase linewidth. Here the contrast between the smallest and largest $\text{Im}[\rho_{12}]$ becomes smaller when $\Omega_p$ or $\gamma_2$ increases. To capture such contrast, we define visibility $V=(R_{max} -R_{min})/(R_{max} +R_{min})$, where $R_{min}$ and $R_{max}$ are the minimum and maximal value of $\text{Im}[\rho_{12}]$. Large visibility means it would be easier to observe the non-uniform profile of the spectrum, and provide strong signals. As shown in Figs.~\ref{fig:absorption(exp)} (c-d), the visibility will decrease due to the sharp decrease of maximum absorption coefficients as $\gamma_2$ and $\Omega_p$ increase. A potential drawback is that smaller $\gamma_2$ leads to slightly wider transparent windows at the AT splitting (Fig. \ref{fig:absorption(exp)}(a)).

\subsection{EIA regime}
In the far detuned case  $|\delta_3|\sim |\delta_2| \gg \delta, \Omega_c$ and $\gamma_2$, the state $|2\rangle$ can be adiabatically eliminated, leading to the EIA scheme (Fig. \ref{fig:level}(c)). This yields an effective three-level Hamiltonian~\cite{Liao2020},
\begin{eqnarray}
H_{\mathrm{eff}} = &&(\delta + \delta_{ac})|3\rangle\langle 3| + \delta_4 |4\rangle\langle 4|\nonumber\\
&& + \frac{\Omega_{\mathrm{eff}}}{2} |1\rangle\langle 3| + \frac{\Omega_m}{2} |3\rangle\langle 4| + H.c.,
\end{eqnarray}
where $\Omega_{\mathrm{eff}} = \Omega_c\Omega_p/2\delta_3$ is an effective Rabi frequency,  and $\delta_{ac}=(\Omega_p^2+\Omega_c^2)/4\delta_3$ is the AC stark shift. When $\Omega_{\mathrm{eff}}\ll \Omega_m$, an analytic solution of the two-photon coherence is obtained
\begin{eqnarray}
\rho_{13}= i \frac{\Gamma_4}{\Omega_m^2+\Gamma_4(\Gamma_3-2i\delta_{ac})}\Omega_{\mathrm{eff}}.
\end{eqnarray}
The maximal absorption occurs when the real part of the denominator vanishes at
\begin{equation}
	\delta_{\pm} = -\frac{1}{2}\left[ \delta_{ac} +\delta_m\pm A(\delta_m)\right],
\end{equation}
with $A(\delta_m)=\sqrt{\Omega_m^2+\gamma_3\gamma_4 +(\delta_{ac} - \delta_m)^2}$. When $\delta_m=0$, we obtain  $\rho_{13}$ at $\delta_{\pm}$
\begin{eqnarray}
	\rho_{13}^{(\pm)} = \frac{iA(0)\pm (\gamma_4 +  i\delta_{ac})}{A(0)(\gamma_4+\gamma_3) \pm \delta_{ac} (\gamma_3-\gamma_4)}\Omega_{\mathrm{eff}},\label{eq:plus}
\end{eqnarray}
which shows that the absorption is asymmetric when $\delta_{ac}\neq 0$.  Note that coherence becomes pure real value,  $\rho_{13}= \frac{2\delta_4\Omega_{\mathrm{eff}}}{\Omega_m^2-4\delta_4(\delta_3+\delta_{ac})}$, when neglecting Rydberg state decay $\gamma_3=\gamma_4=0$.

When $\delta_m=0$ and $\Omega_m\gg \delta_{ac}$, the coherence is approximately given by,
\begin{eqnarray}
\rho_{13} \approx -\frac{\Omega_{\mathrm{eff}}}{2}\left[\frac{1}{2\delta + \Omega_m  + i\gamma_3 }+\frac{1}{2\delta - \Omega_m  + i\gamma_3 }\right], \label{eq:symmetry}
\end{eqnarray}
It contains two equal-width (width $\gamma_3$) Lorentzian peaks shifted by $\pm \Omega_m/2$ from the two-photon resonance frequency. The two absorption peaks of the coherence can be used to sense MW fields~\cite{Urvoy2013} whose splitting equals to $\Omega_m$, as been demonstrated in a recent experiment~\cite{Liao2020}.

When $\delta = \pm \Omega_m/2$, the saturation parameter is obtained,
\begin{eqnarray}
s = \left(\frac{\Omega_{\mathrm{eff}}}{2\gamma_3}\right)^2=\left(\frac{\Omega_{c}\Omega_{p}}{4\gamma_3\delta_3}\right)^2.
\end{eqnarray}
Compared to the EIT regime, the saturation parameter in the EIA regime will rely on the Rydberg state decay rate. For given Rydberg states, strong absorption can be obtained by using large $\Omega_c$ and $\Omega_p$, and small $\delta_3$. Note that the latter means that $\gamma_2$ needs to be small, as the adiabatic elimination requires $\delta_3 \gg \gamma_2$.

\begin{figure}[h]
	\includegraphics[width=0.48\textwidth]{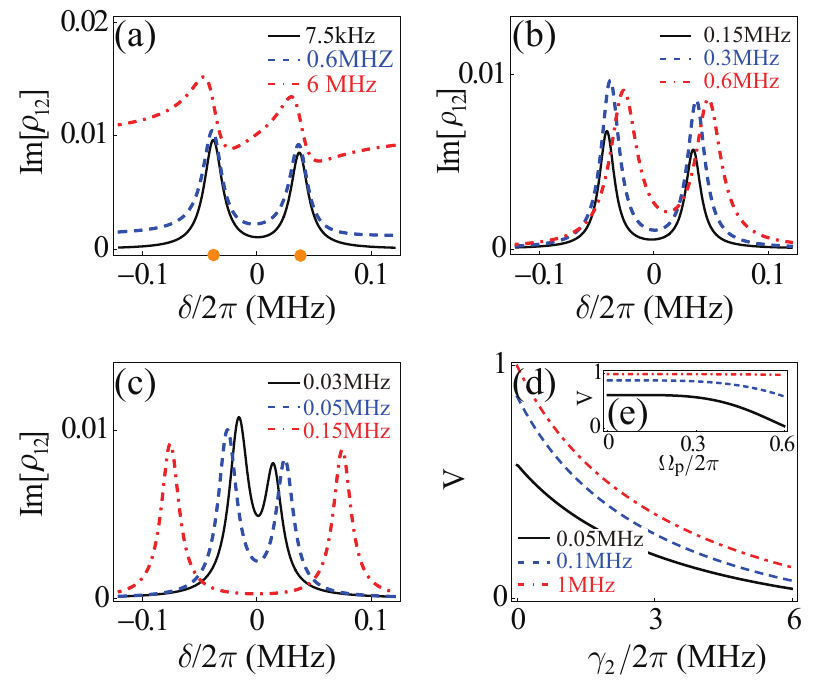}
	\caption{(a) The absorption spectrum with $\gamma_2/2\pi =  7.5$ kHz (black solid line), $0.6$ MHz (blue dashed line) and $6$ MHz (red dot-dashed line) in the far detuning case ($\delta_3/2\pi= 6$MHz). The other parameters are $\gamma_3/2\pi = \gamma_4/2\pi = 15$ kHz, $\Omega_p/2\pi =0.3$ MHz, $\Omega_c/2\pi = 0.45$ MHz, $\Omega_m = 5\gamma_3$, and  $\delta_\mathrm{m} = 0$. The filled circles in the axis give the locations of the maximal absorption. (b) The absorption spectrum with $\Omega_p/2\pi =  0.15$ MHz (black solid line), $ 0.3$ MHz (blue dashed line) and $0.6$ MHz (red dot-dashed line). Here $\gamma_2/2\pi =  7.5$kHz and the other parameters are the same with (a). (c) The absorption spectrum with $\Omega_m/2\pi =  0.03$ MHz (black solid line), $ 0.05$ MHz (blue dashed line) and $ 0.15$ MHz (red dot-dashed line). (d) Visibility as function of $\gamma_2$ with $\Omega_m/2\pi = 0.05$ MHz (black solid line), 0.1 MHz (blue dashed line) and 1 MHz (the red dot-dashed line). (e) Visibility as function of $\Omega_p$ with $\Omega_m/2\pi = 0.05$ MHz ( black solid line), 0.1 MHz (blue dashed line) and 1 MHz (red dot-dashed line). The other parameters in (c-e) are the same with (b).}
	\label{fig:detuning}
\end{figure}

In Fig. \ref{fig:detuning} we show the absorption spectrum obtained  by numerically solving the original master equation (2). The absorption peaks agree with the analytical result, especially when $\gamma_2$ is small. Though $\delta_{\pm}$ is  independent of $\gamma_2$, the width of the absorption peak increases with increasing $\gamma_2$. $\delta_{\pm}$ is not symmetric and depends non-negligibly on Rabi frequencies of the probe and MW fields. In Figs.~\ref{fig:detuning}(b), the location of the peaks shifts to the positive $\delta$ with increasing $\Omega_p$, due to the Stark shift, though the distance between the peaks is unchanged. In Fig.~\ref{fig:detuning}(c), one can observe clearly when MW filed is weak, $\Omega_m \leq \delta_{ac}$, the two absorption peaks are asymmetric, as shown in Eq.~(\ref{eq:plus}). When $\Omega_m$ is strong enough, $\Omega_m \gg \delta_{ac}$, the symmetry of the two absorption is nearly restored, which is consistent with Eq.~(\ref{eq:symmetry}). One important figure of the merit is the contrast between the height of the peaks and the lowest value between the two peaks. The resulting visibility is plotted in Figs.~\ref{fig:detuning} (d-e). Similar with the EIT case, the visibility increase with increasing $\Omega_m$ (decreasing $\gamma_2$) with fixed $\gamma_2$ ($\Omega_m$). Moreover, when $\Omega_p >\Omega_c$ the visibility decreases dramatically with increasing $\Omega_p$.

\begin{figure}[h]
\includegraphics[width=0.48\textwidth]{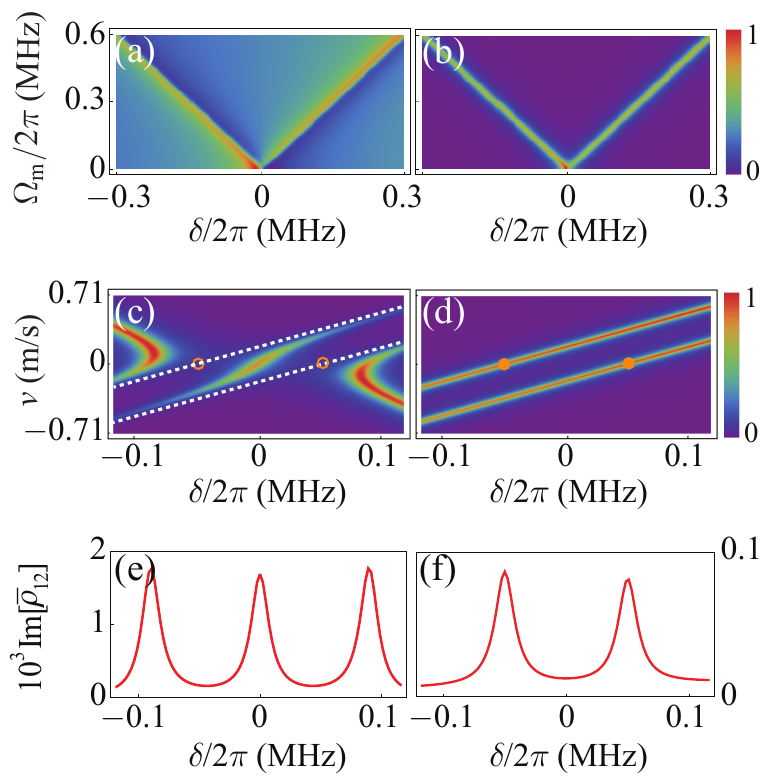}
\caption{(a-b) The absorption spectrum with (a) $\gamma_2 =  2\pi \times 6$ MHz, and (b) $\gamma_2 = 2\pi \times 7.5 $ kHz. The other parameters are $\gamma_3 = \gamma_4 =2\pi \times 15 $ kHz, $\Omega_\mathrm{p} = 2\pi\times 0.3$ MHz, and $\Omega_\mathrm{c} = 0.45$ MHz, $\delta_\mathrm{m} = 0, \delta_3=2\pi\times 6$ MHz. (c-d) Absorption (normalized) vs atom velocity and detuning $\delta$ with $T=300 $K, $\gamma_2 = 2\pi\times7.5$ kHz, $\gamma_3 = \gamma_4 = 2\pi\times 15$ kHz and $\Omega_m =2\pi\times0.1$ MHz. The dashed lines in (c) show the absorption dips are shifted. The open (filled) circles indicate the absorption dips (peaks) locations when the atoms do not move, i.e., $v=0$. The other parameters for (c) EIT regime: $\delta_3=0, \Omega_p= 2\pi\times 15$ kHz, $\Omega_c= 2\pi\times 150$ kHz, (d) EIA regime: $\delta_3= 2\pi\times 6$ MHz, $\Omega_p= 2\pi\times 0.3$ MHz, $\Omega_c= 2\pi\times 0.45$ MHz. (e-f) The corresponding Doppler averaged absorption spectrum in the (e) EIT  and (f) EIA regime.}
\label{fig:detuning2d}
\end{figure}

In Figs. \ref{fig:detuning2d}(a-b), we compare the absorption spectrum when $\gamma_2 = 2\pi\times 6$ MHz (a) and $\gamma_2=2\pi\times 7.5\ \text{kHz}$ (b), respectively. The resulting absorption spectrum is asymmetric in Fig.~\ref{fig:detuning2d}(a), as the large detuning condition is not fully met. The absorption peaks sit on a non-negligible background such that the contrast is low. To meet the EIA condition, even larger $\delta_3$ is needed, which would lead to weaker absorption~\cite{Liao2020}. On the contrary, we can balance the large detuning condition and high saturation to achieve high contrast when $\gamma_2$ is small. As shown in Fig. \ref{fig:detuning2d}(b), the absorption spectrum becomes symmetric and the distance between the EIA peaks is $\Omega_m$. The above results show that the EIA spectrum of Sr triplet states is robust. The high visibility might be beneficial to sensing weak MW fields.

\subsection{Finite temperature effects}
In this section, we briefly discuss the Doppler effect due to finite temperature $T>0$ of the atomic gas. We will show that the main conclusions are still valid, despite that the (absorption/transmission) signal becomes weaker. We assume that the probe and control light counter propagate, where the Doppler effect is integrated into the detuning through $\delta_p' = \delta_2 + (2\pi/\lambda_p) v$, $\delta_c' = \delta_3 - (2\pi/\lambda_c) v$, with $v$ the atomic velocity~\cite{Holloway2017, Kwak2016}. Doppler shifts of the MW field are small and can be ignored. We average steady state solution of $\rho_{12}$ over the atomic velocity distribution that is found in a vapor cell at temperature $T$ with $\bar{\rho}_{12}=\int dv f(v)\rho_{12}(\delta_p',\delta_c')$. Here $f(v)=1/(\sqrt{\pi}u)\text{exp}[-(v/u)^2]$ is the Maxwell-Boltzmann velocity distribution \cite{Holloway2017, Bai2020PRL} with $u=\sqrt{2k_BT/m}$ the root mean square atom velocity at temperature $T$ and $m$ the mass of the atom.

The absorption spectrum of the atoms with different velocities and the Doppler averaged  absorption spectrum in the EIT regime and EIA regime are shown in Fig.~\ref{fig:detuning2d}(c-f). In Fig.~\ref{fig:detuning2d}(c), the absorption peak in the center $\delta=0$ is shifted by $2\pi/\lambda_p v$ and represents the one-photon resonance signal of the probe beam.  The two absorption minima around $\delta=\pm{\Omega_m}/2$ (the open circles, at zero temperature) are shifted by the Doppler shift $(2\pi/\lambda_p - 2\pi/\lambda_c) v$ of the two-photon resonance (the dashed lines in Fig.~\ref{fig:detuning2d}(c)) due to the wavelength difference between the probe and control fields. As a result, the contribution from each velocity class strongly reduces the visibility of probe absorption spectrum compared to the Doppler-free case, as depicted in Figs.~\ref{fig:detuning2d}(e). However in the EIA regime, as shown in Fig.~\ref{fig:detuning2d}(d), the two absorption peaks are shifted by the Doppler shift of the two-photon resonance to be $(2\pi/\lambda_p - 2\pi/\lambda_c) v$ and the one-photon Doppler shift is absence because of the large one-photon detuning. The above analysis shows that the EIA feature persists even in the presence of Doppler effects (Figs.~\ref{fig:detuning2d}(f)).

\begin{figure}
	\includegraphics[width=0.48\textwidth]{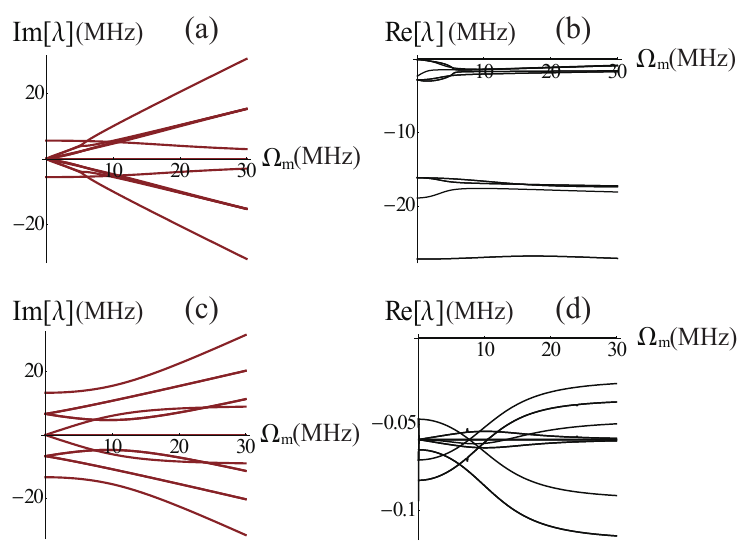}
	\caption{The imaginary and real part of the spectrum of the Liouville operator $\mathcal L$  when (a, b) $\gamma_2 = 2\pi\times 6$ MHz  and (c-d) $\gamma_2 = 2\pi \times 7.5$ kHz. The other parameters are $\gamma_3 = \gamma_4 = 2\pi \times 15$ kHz, $\Omega_p = \Omega_{c}=2\pi\times1.5$ MHz, and $\delta_{2} = \delta_3 = \delta_{m} = 0$.}
	\label{fig:eigenvalue}
\end{figure}
\section{Resonance fluorescence}
\subsection{Time scales in the EIT and EIA schemes}

So far our analysis has been focused on the stationary coherence. When $\gamma_2$ is small, one would expect that the system takes longer time to reach the stationary state. The response time of the system depends on a number of parameters~\cite{macieszczak_metastable_2017,zhang_fast-responding_2018,guo_transient_2020}.
Here we investigate the time scale of the dynamics in the EIT and EIA regime through analyzing the master equation. We calculate
eigenvalues $\lambda_k$ of the Liouville operator $\mathcal L$~\cite{Bienert2004},
\begin{eqnarray}
	\mathcal L R_k &=& \lambda_k R_k,\\
	L_k \mathcal L &=& \lambda_k L_k,
\end{eqnarray}
with eigenvalues $\lambda_k, (k=1,2,...)$, left and right eigenvectors $L_k$ and $R_k$, respectively. The orthogonality and completeness of the eigenvectors  is defined with respect of the trace, such that $\mathrm{Tr}\{L_k R_{k'}\} = \delta_{k,k'}$. Eigenvalue $\lambda_1 =0$ and the corresponding eigenvector give the stationary state of the master equation.

If the MW field and the decay rate  are small, $\Omega_m,\gamma_2 \ll \Omega_{p},\Omega_c$, the imaginary part of the eigenvalues can be obtained approximately. The first two sets of solutions are  $\mathrm{Im}[\lambda]\simeq\{0 ,\pm\sqrt{\Omega_{p}^2 + \Omega_{c}^2}\}$ independent of $\Omega_m$. The other two sets of solutions to $\mathrm{Im}[\lambda]$ are
\begin{equation}
 \{\pm 9 \sqrt{\frac{\gamma_2}{\Omega_{c}}} \Omega_m,\frac{\pm\sqrt{\Omega_{p}^2 + \Omega_{c}^2}}{2}\pm 6 \sqrt{\frac{\gamma_2}{\Omega_{c}}} \Omega_m\}.
\end{equation}
The approximate solutions show that $\text{Im}[\lambda_k]$ that are between 0 and the maximum (minimum) will be linearly changing with $\Omega_m$. In Fig.~\ref{fig:eigenvalue}, the numerical spectrum of operator $\mathcal{L}$ as function of $\Omega_{m}$ for $\gamma_2=2\pi\times 6$ MHz (a-b) and $\gamma_2 = 2\pi\times 7.5$ kHz (c-d) are shown. The linear dependence can be seen in the figure when $\Omega_m$ is small.

\begin{figure}
	\includegraphics[width=0.48\textwidth]{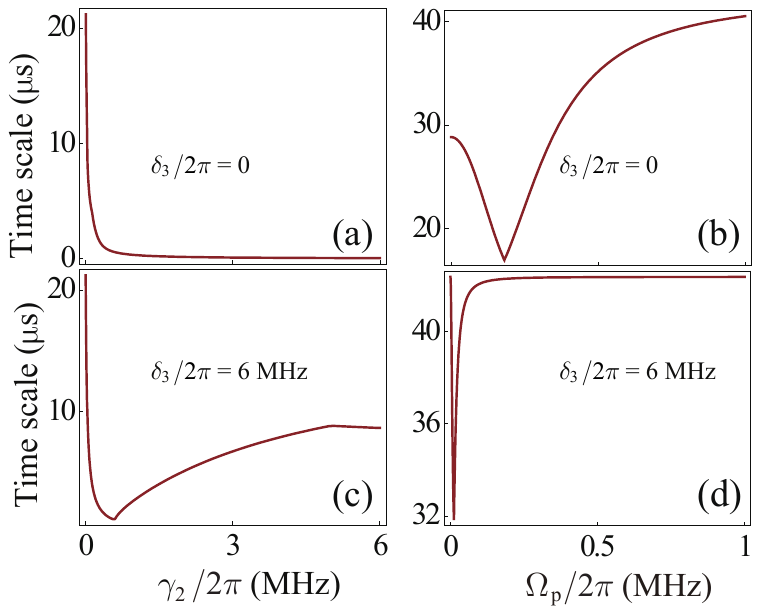}
	\caption{Time scale in the EIT (a-b) and EIA (c-d) regime. We consider (a) $\Omega_p=2\pi\times 7.5$ kHz and (b) $\gamma_2=2\pi\times 7.5$ kHz with $\gamma_3 = \gamma_4 = 2\pi\times 15$ kHz, $\delta_2=\delta_3=\delta_m=0$, $\Omega_c=2\pi\times 75$ kHz and $\Omega_m=2\pi\times 0.16$ MHz. In the EIA regime, we consider (c) $\Omega_p=2\pi\times 0.3$ MHz and (d) $\gamma_2=2\pi\times 7.5$ kHz with $\delta_3=2\pi\times 6$ MHz. The other parameters are $\gamma_3 = \gamma_4 = 2\pi\times 15$ kHz, $\delta_2=\delta_m=0$, $\Omega_c=2\pi\times 0.45$ MHz and $\Omega_m=2\pi\times 0.16$ MHz.}
	\label{fig:timescale}
\end{figure}

$|\mathrm{Re}[\lambda_k]|$ gives the relaxation rate of the $k$-th mode. Inverse of the real part of the second largest eigenvalue, $1/|\mathrm{Re}[\lambda_2]|$ gives the longest time required for the system to evolve to a stationary state~\cite{Macieszczak2016a}. We plot the longest time scale in EIT ($\delta_3=0$ , Figs.~\ref{fig:timescale}(a-b)) and EIA ($\delta_3=2\pi\times 6$ MHz Figs.~\ref{fig:timescale}(c-d)) regim, respectively. For large $\gamma_2$, it takes shorter times to reach steady state in EIA than EIT regime. When $\gamma_2$ is small, the timescale between EIT and EIA scheme is similar.  As shown in Figs.~\ref{fig:timescale}(b) and (d), the time scale no longer changes monotonously with $\Omega_p$. A sharp dip appears in the time scale due to the degeneracy of $\mathrm{Re}[\lambda_2]$ and $\mathrm{Re}[\lambda_3]$. The above analysis shows that when $\gamma_2$ is small, it could take significant amount of time to reach the steady state. Such slow time scale could be probed, for example, by measuring the time-dependent probe light transmission~\cite{zhang_fast-responding_2018}.

\subsection{Spectrum of resonance fluorescence}
In the rest of the work, we investigate the spectrum of resonance fluorescence of Sr atoms. We will identify a linear relation between the spectrum and MW field, which could provide a new way to measure weak MW fields. The power spectrum is given by the transform of the optical field autocorrelation function~\cite{Steck2018},
\begin{equation}
S(\omega,t) = \mathrm{Re} \int_0^{\infty} d \tau e ^{-i\omega \tau} \langle \sigma_{21}(t) \sigma_{12}(\tau+t)\rangle.
\end{equation}
Using quantum regression theorem,  the two-time correlation function can be calculated,
\begin{equation}
\langle \sigma_{21}(t) \sigma_{12}(t+\tau)\rangle = \mathrm{Tr} [\sigma_{12}\Lambda(t+\tau,t)],
\end{equation}
where $\Lambda(t+\tau,t)$ satisfies the evolution equation
\begin{equation}
\partial_{\tau}\Lambda(t+\tau,t) = \mathcal{L}\Lambda(t+\tau,t),
\end{equation}
and the initial condition is
\begin{equation}
\Lambda(t,t) = \rho(t)\sigma_{21}.
\end{equation}
One can find that $\Lambda(t+\tau,t) = e^{\mathcal{L} \tau}\Lambda(t,t)$, so that
\begin{eqnarray}
S(\omega,t)&=&\mathrm{Re} \int_0^{\infty} d \tau e ^{-i\omega \tau} \mathrm{Tr}[\sigma_{12} \Lambda(t+\tau,t)]\nonumber\\
&=&\mathrm{Re}[\frac{1}{i\omega-\mathcal{L}}\sum_k  \mathrm{Tr}[\sigma_{12} e^{\tau \lambda_k}c_k R_k \sigma_{21}]],
\end{eqnarray}
where $c_k = \mathrm{Tr}[L_k\rho_{in}]$.

When $t\rightarrow\infty$, $\lim_{t\rightarrow\infty}\Lambda(t,t) = \rho_{ss}\sigma_{21} = \Lambda(0)$. The corresponds stationary spectrum can be obtained,
\begin{eqnarray}
S(\omega)&=&\mathrm{Re} \int_0^{\infty} d \tau e ^{-i\omega \tau} \mathrm{Tr}[\sigma_{12} \Lambda(\tau)]\nonumber\\
&=&\mathrm{Re}[\sum_k \frac{1}{i\omega - \lambda_k} \mathrm{Tr}[\sigma_{12} c_k R_k \sigma_{21}]],
\end{eqnarray}
which indicates the spectrum of resonance fluorescence is the sum of Lorentz profile, centered at the imaginary part of the eigenvalues of $\mathcal L$, with width given by the real part of $\lambda_k$. The fluorescence spectrum can also be understood by analyzing eigenenergies of the dressed states~\cite{Wang2009,Wang2011, Tian2012}.

\begin{figure}[h]
\includegraphics[width=0.48\textwidth]{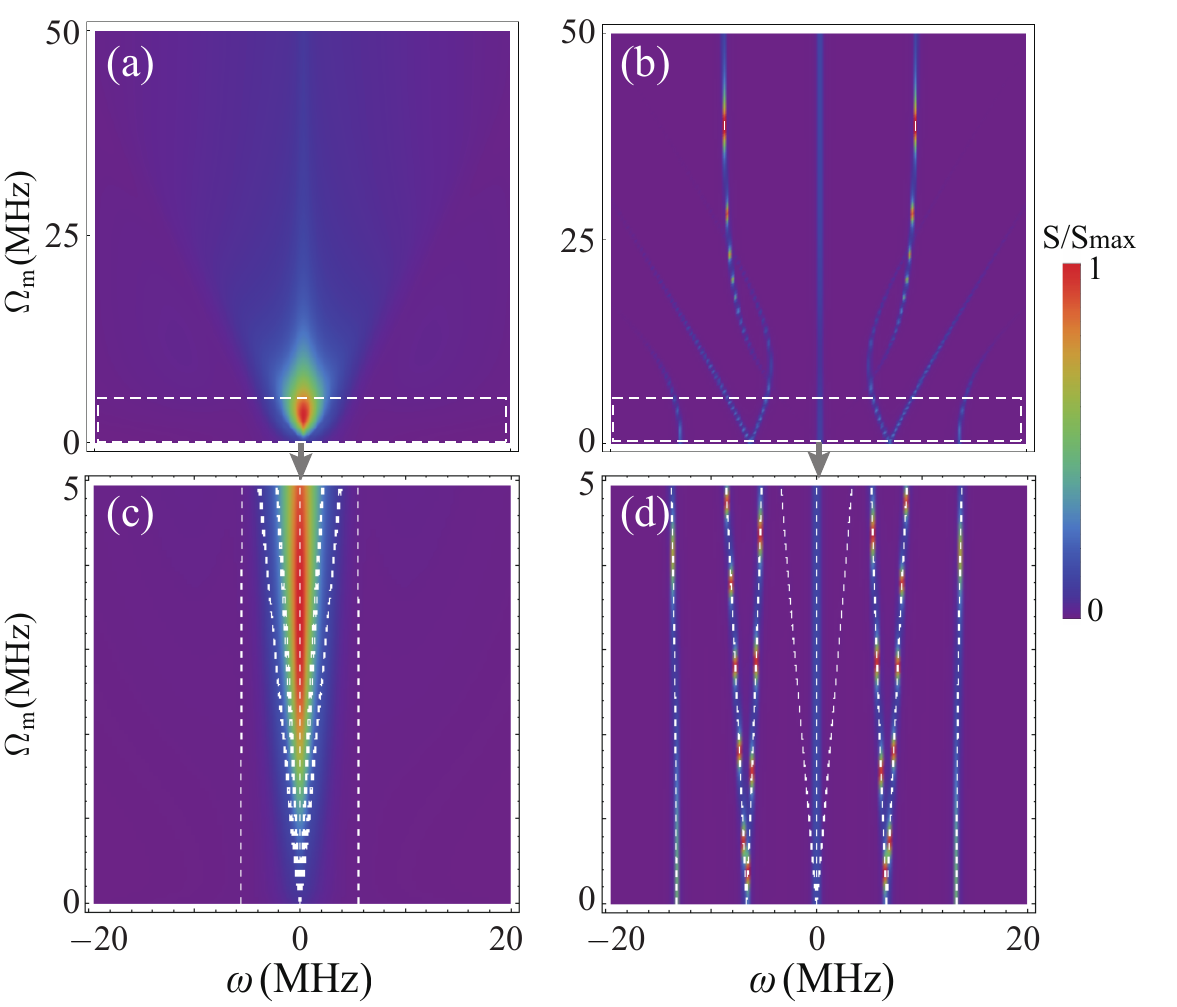}
\caption{The spectrum of resonance fluorescence as function of $\Omega_\mathrm{m}$ with (a) $\gamma_2 = 2\pi\times 6$ MHz, (b) $\gamma_2 = 2\pi\times7.5 $ kHz. Panels (c) and (d) show the enlarged region marked by the white box in panel (a) and (b). The dashed lines indicate imaginary parts of $\lambda_k$. The other parameters are the same with Fig. \ref{fig:eigenvalue}.}
\label{fig:florance2D}
\end{figure}

In the stationary state, the peaks acquires wider linewidth when $\gamma_{2}$ is large (see Fig.~\ref{fig:eigenvalue}(b)). The wide linewidth will suppress peaks for $\omega\neq 0$ (see Figs.~\ref{fig:florance2D} (a, c)). When $\gamma_2$ is smaller than $\Omega_p(c)$, the system is in the so-called strongly driven regime. A well studied example in this regime is the so-called Mollow triplet of two-level atoms~\cite{PhysRev.188.1969}. Multiple peaks can be observed in spectrum of the resonance fluorescence, and the distance between the peaks near the center peak increases linearly with increasing $\Omega_m$ for weak MW fields, as shown in Figs.~\ref{fig:florance2D}(b, d). These peaks have particularly narrow linewidth, as the real parts of $\lambda_k$ are small (see Fig.~\ref{fig:eigenvalue}(d)). The linear dependence on $\Omega_m$ potentially allows us to identify the MW field Rabi frequency (e.g. when the strength of the MW field drifts slowly with time) by measuring the distance between these two peaks, opening perspectives to develop Rydberg MW electrometry based on Sr atoms.

\begin{figure}[h]
\includegraphics[width=0.43\textwidth]{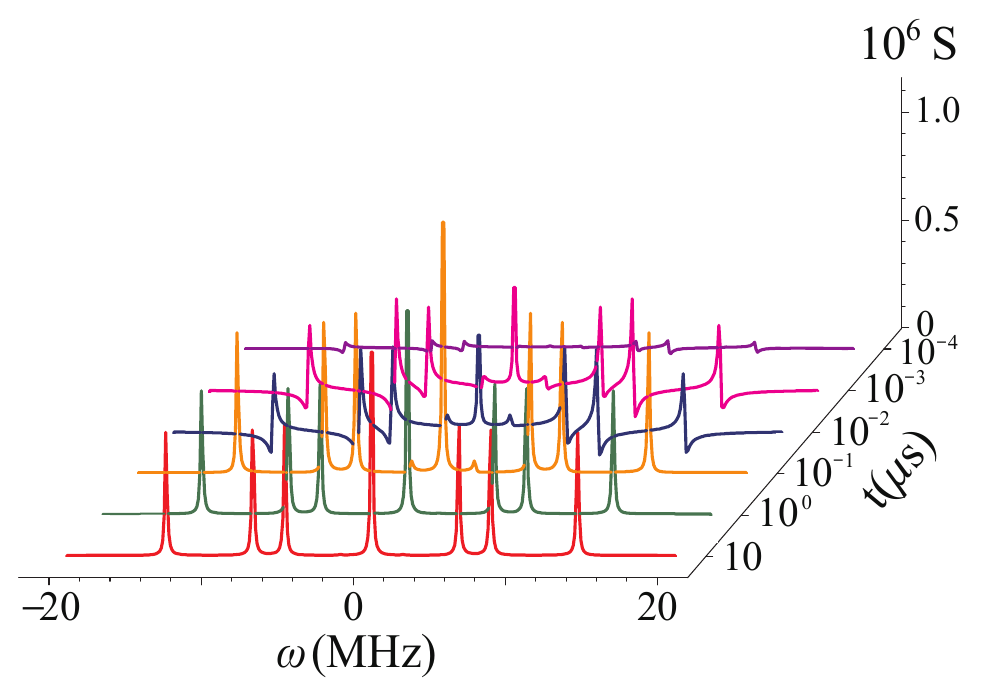}
\caption{Time dependent fluorescence spectrum for $\gamma_2 = 2\pi\times 7.5$ kHz. Other parameters are $\gamma_3 = \gamma_4 = 2\pi\times 15$ kHz, $\Omega_{m} \approx  2\pi\times 0.48$ MHz, and $\Omega_{c} = \Omega_{p}=2\pi\times 1.5$ MHz. }
\label{fig:florance(t)}
\end{figure}

Finally, we address the time dependent fluorescence spectrum, shown in Fig. \ref{fig:florance(t)}. Initially, heights of the fluorescence peaks are low.  These peaks increase with increasing interaction time. The spectrum approaches to the stationary distribution when $t>10\,\mu$s. Here the time scale in the dynamical evolution of the spectrum is largely determined by $1/\gamma_2$, consistent with the analysis of the eigenvalues of the Liouville operator. These time scale is much smaller than the motional dephasing in cold and ultracold atom gases, and hence could be measured in current experiments.

\section{Conclusion}
We have studied optical responses of Sr triplet optical transition and showed that weak MW field can be sensed through combined using EIT and AT splitting, as well as EIA. The significantly small decay rate in the triplet state affects the transparency spectrum in the Rydberg EIT. The stationary transmission, compared with alkali atoms, becomes weaker. Narrow and strong absorption peaks are found neighboring to the weakened transparent window, leading to apparent differences between the transparent and absorption strength (Fig.~\ref{fig:absorption(exp)}). In the EIA regime, the maximal absorption is increased from the background absorption.  The distance of the transparent windows (EIT) and the absorption peaks (EIA), given by $\Omega_m$, can be as small as the decay rate $\gamma_2$ of the intermediate state. The high contrast in both approaches is beneficial when sensing MW field with Sr Rydberg states.  Moreover, we have examined resonance fluorescence of the lower transition in the four-level system. Mathematically it is found that the location of the florescence peaks is determined by the imaginary parts of the Liouville operator. We have shown that the distance between neighboring peaks is linearly proportional to the MW field Rabi frequency when $\Omega_m\sim \gamma_2\ll \Omega_c,\Omega_m$. Such dependence is not visible in the case of alkali atoms due to that the decay rate in the respective  state is large. one might be able to measure weak MW fields coupled to triplet Rydberg states with Sr atoms.  In the EIT-AT and EIA method, one measures directly transmission and absorption of the probe light. In resonance fluorescence, photons scattered by the atoms are collected and measured. This method provides a unique opportunity to sense MW fields when directly measuring the transmission or absorption is not possible, or difficult to achieve.

\begin{acknowledgments}
We thank insightful discussions with Dr. Peng Xu and Prof. Hui Yan. This work was supported by the National Nature Science Foundation of China (Grants No. 12074433, 11871472, 11874004, 11204019, 12174448), National Basic Research Program of China (Grant No. 2016YFA0301903), Science Foundation of Education Department of Jilin Province (JJKH20200557KJ), Nature Science Foundation of Science and Technology Department of Jilin Province (20210101411JC).
\end{acknowledgments}

\bibliography{bibfile}

\end{document}